s
\documentclass[12pt]{iopart}
\usepackage{graphicx}

\usepackage{iopams}
\usepackage{amssymb}
\usepackage{color}

\begin{document}

\title[Quantumness domains and squeezed states]{New quantumness domains through generalized squeezed states}

\author{F A Raffa$^1$, M Rasetti$^{1,2}$, M Genovese$^3$}

\address{$^1$ Politecnico di Torino, Dipartimento Scienza Applicata e Tecnologia, Corso Duca degli Abruzzi 24, 10129 Torino, I}
\address{$^2$ ISI Foundation, Institute for Scientific Interchange, Via Chisola 5, 10126 Torino, I}
\address{$^3$ INRiM,  Strada delle Cacce 91, 10135 Torino, I}
\ead{francesco.raffa@polito.it}
\vspace{10pt}
\begin{indented}
\item[]August 2019
\end{indented}

\begin{abstract}
Current definitions of both squeezing operator and squeezed vacuum state are critically examined on the grounds of consistency with the underlying su(1,1) algebraic structure. Accordingly, the generalized coherent states for su(1,1) in its Schwinger  two-photon realization are proposed as squeezed states. The physical implication of this assumption is that two additional degrees of freedom become available for the control of quantum optical systems. The resulting physical predictions are evaluated in terms of quadrature squeezing and photon statistics, while the application to a Mach-Zehnder interferometer is discussed to show the emergence of nonclassical regions, characterized by negative values of Mandel's parameter, which cannot be anticipated by the current formulation, and then outline future possible use in quantum technologies.
%
\vspace{2pc}
\noindent\\ Keywords: generalized coherent states, squeezed vacuum, photon statistics, quantumness, Mach-Zender interferometer
\end{abstract}
%
\submitto{\JPA}
%
\maketitle
%
%

\section{Introduction}\label{INTRO}
Let's proceed first to a brief review of some definitions concerning generalized coherent states and their relation to groups and algebras. Generalized coherent states $|\zeta\rangle$ for a given algebra $A$ are obtained through a generic unitary operator $U$ $=$ $e^{\hat{g}}$ $\in$ $G$, where $G$ is the group associated with $A$, and $\hat{g}$ is the most general anti-hermitian element of $A$, $\hat{g}^\dagger$ $=$ $- \hat{g}$. One defines $|\zeta\rangle$ $=$ $U |\omega\rangle$, where $|\omega\rangle$ denotes the relevant `‹vacuum' state, i.e., the state annihilated by the lowering operators of algebra $A$ (in mathematical jargon the highest weight vector) and $\zeta$ is the set of $c$-numbers, which parametrize $\hat{g}$ and $|\omega\rangle$. This theoretical framework was first devised in \cite{PEREart}-\cite{MARIO}; comprehensive reviews as well as applications in many fields of fundamental physics can be found in \cite{PEREbook}-\cite{BOHM}.

In the current formulation of quantum optics squeezing, squeezed vacuum states of light are only a subset of  $|\zeta\rangle$ for $G$ $\equiv$ SU(1,1), because $\hat{g}$ is not the most general element of $A$ $\equiv$ su(1,1), generated by the set of operators $\{K_0 \textrm{(Cartan)}, K_- \textrm{(lowering)}, K_+ = K_-^\dagger\textrm{(raising)} \}$ in its Schwinger two-boson realization, and the vacuum state $|\omega\rangle$ is identified with the physical vacuum $|0\rangle$, which is not the highest weight vector of $A$.

The importance of such conventional coherent states cannot be underestimated due to their several applications, mainly based on their sub-Poissonian statistics. Nowadays they are crucially related to the high sensitivity measurements required in quantum-enhanced metrology, particularly for the detection of gravitational waves \cite{KENZIE}-\cite{OELK} and quantum gravity tests \cite{GENOVESE}-\cite{PRADY}, as well as photoelectric detection \cite{VAHL}, absorption measurements \cite{PARIS} and the analysis of the Casimir effect \cite{PAN}. They play a crucial role also in quantum cryptography and quantum information as reported in \cite{FRANCE}, where the efficiency of single-photon sources is investigated, and in the technology of photon cutting, which aims at improving the energy conversion efficiency in optical materials \cite{JONG}. Because of their remarkable properties, they are plenty of attention in textbooks \cite{MANDEL}-\cite{FOX} and in review papers \cite{LOKN}-\cite{GAZE}. An important role of squeezed states, on which we focus our attention in this paper, is their feature of quantumness estimated in quantum optics by various indicators such as the sign of the Mandel parameter $Q$ \cite{QM}.

In this work we argue in favor of the adoption as quantum optical squeezed states of the generalized coherent states of su(1,1) and explore the ensuing effects. This implies including $K_0$ in $\hat{g}$, and assuming as vacuum state a linear combination of the ordinary vacuum state $|0\rangle$ and the one photon state $|1\rangle$, which is the most general state annihilated by $K_-$. The main feature of this approach consists thus in the higher dimension of the space of parameters. Its physical reach is mostly contained in the squeezing properties of the quadratures and in the photon statistics. This is shown in the specific example of a Mach-Zehnder interferometer in which the generalized squeezed state is sent through one of the input ports. In this application novel quantumness regions of the system appear, induced by the emergence of sub-Poissonian statistics, identified just by negative values of $Q$.

The paper is organized as follows. The definitions and the relevant variances of the quadratures are reviewed  for the conventional approach in Sect. \ref{TRUAX} and for the generalized approach in Sect. \ref{RR}. Sec. \ref{SQ} is devoted to the analysis of squeezing, particularly to the identification of which quadrature is actually squeezed. In Sect. \ref{PHOTO} the photon probability distributions are discussed. In Sect. \ref{MANDMZ} the working example of the Mach-Zehnder interferometer is considered with special attention to the occurrence of new sub-Poissonian regions. A few concluding remarks are given in Sect. \ref{END}.

\section{Conventional formulation}\label{TRUAX}
In the conventional approach to quantum optical squeezing, the unitary squeezing operator $S(\tau)$ is the exponential of an anti-hermitian linear combination of two operators  $K_+$ and $K_-$ $=$ $K_+^\dagger$,
\[
S(\tau) = e^{\tau K_+ - \bar{\tau} K_-} \, ,
\]
where $\tau \in \mathbb{C}$, while $K_+$, $K_-$ are defined in terms of $a$ and $a^\dagger$, the harmonic oscillator annihilation and creation operators,
\begin{equation}\label{K+-}
K_- = \frac{1}{2} a^2 \; , \; K_+ = \frac{1}{2} a^{\dagger 2} \, .
\end{equation}
The introduction of operator $K_0$,
\begin{equation}\label{K0}
K_0 = \frac{1}{2} \left(\hat{n} + \frac{1}{2}\right) \, ,
\end{equation}
where $\hat{n}$ $=$ $a^\dagger a$ is  the harmonic oscillator number operator, closes an algebra. Indeed, the set $\{K_0, K_+, K_-\}$, characterized by commutation relations $[K_0,K_\pm]$ $=$ $\pm K_\pm$, $[K_+,K_-]$ $=$ $- 2 K_0$, generates the algebra su(1,1) in its Schwinger two-boson realization. $K_+$ and $K_-$ are the raising and lowering operators, respectively, of the algebra; therefore $S(\tau)$ is an element of the group SU(1,1) in this realization. The squeezed vacuum state is then obtained through application of $S(\tau)$ to the physical vacuum state $|0\rangle$ annihilated by $K_-$,
\[
|\tau\rangle_0 = S(\tau) |0\rangle \, .
\]

We focus on variances $\Delta^2(\bullet)$ $\doteq$ $\langle \bullet^2 \rangle - \langle \bullet \rangle^2$, where $\langle \bullet \rangle$ $=$ $_0\langle \tau| \bullet |\tau\rangle_0$ $=$ $\langle 0| S^\dagger (\tau) \bullet S(\tau) |0\rangle$ (note that all expectations are thus taken in state $|0\rangle$) for the quadratures,
\begin{equation}\label{QUAD}
q = \frac{1}{\sqrt{2}} (a^\dagger + a) \; , \; p = \frac{i}{\sqrt{2}} (a^\dagger - a) \, ,
\end{equation}
with
\begin{eqnarray*}
q^2  &=&\left( \hat{n} + \frac{1}{2} \right) + \frac{1}{2} a^{\dagger 2} + \frac{1}{2} a^2 = 2 K_0 + K_+ + K_- \, , \\
p^2  &=&\left( \hat{n} + \frac{1}{2} \right) - \frac{1}{2} a^{\dagger 2} - \frac{1}{2} a^2 = 2 K_0 - K_+ - K_- \, .
\end{eqnarray*}

In actual calculations a disentangled version of $S(\tau)$ is required. In normal order form one writes
\[
S(\tau) = e^{t_+ K_+} e^{t_0 K_0} e^{t_- K_-} \, ,
\]
where coefficients $t_+$, $t_0$, $t_-$, which depend on $\tau$, can be obtained through the method of Truax \cite{TRUAX}. With the upper bar denoting complex conjugation one has\footnote{For notational convenience, the dependence of the disentanglement coefficients (\ref{T+h})-(\ref{T0h}) as well as of (\ref{P+h})-(\ref{P0h}) on the parameters appearing in the squeezing operators is omitted everywhere.}
\begin{eqnarray}
t_+ &=& \frac{\tau}{|\tau|} \tanh |\tau| \, , \, t_- = - \bar{t}_+  \, , \label{T+h} \\
t_0 &=& - 2 \ln (\cosh |\tau|) \, . \label{T0h}
\end{eqnarray}
Transformation
\[
S^\dagger (\tau) a^\dagger S(\tau) = \cosh|\tau| \, a^\dagger + \frac{\bar{\tau}}{|\tau|} \sinh|\tau| \, a \, ,
\]
is sufficient to evaluate the transformations of all relevant operators. The variances $\Delta^2(q)_0$, $\Delta^2(p )_0$ of quadratures (\ref{QUAD}) with respect to $|\tau\rangle_0$, setting $\tau$ $=$ $|\tau| e^{i \Phi_\tau}$, turn out to be
\begin{eqnarray}
\Delta^2(q)_0 &=& \frac{1}{2} + \sinh^2|\tau| + \sinh |\tau| \cosh |\tau| \cos \Phi_\tau \, , \label{VARq0} \\
\Delta^2(p )_0 &=& \frac{1}{2} + \sinh^2|\tau| - \sinh |\tau| \cosh |\tau| \cos \Phi_\tau \, . \label{VARp0}
\end{eqnarray}

\section{Generalized formulation}\label{RR}
We construct now the generalized squeezed states through the following steps:
\begin{enumerate}
\item The unitary squeezing operator is assumed to be the most general element of SU(1,1),
\begin{equation}\label{U}
U(\alpha,\tau) = e^{i \alpha K_0 + \tau K_+ - \bar{\tau} K_-} \, ,
\end{equation}
where $\alpha \in \mathbb{R}$.
\item The generalized `‹vacuum' state $|\omega\rangle $ is the normalized linear combination of states $|0\rangle$ and $|1\rangle$,
\begin{equation}\label{OMEGA}
|\omega\rangle = \cos \vartheta |0\rangle + \sin \vartheta |1\rangle \, ,
\end{equation}
 parametrized by a single real angle $\vartheta$, which manifestly satisfies the requirement $K_- |\omega\rangle$ $=$ 0.
\item By construction the generalized squeezed vacuum,
\begin{equation}\label{GEN}
|\alpha, \tau, \vartheta\rangle = U(\alpha,\tau) |\omega\rangle \, ,
\end{equation}
coincides with the generalized coherent states of  su(1,1) in the Schwinger realization, where $\{\alpha, \tau, \vartheta\}$ is the set referred to as $\zeta$ in the Introduction.
\end{enumerate}
The normal order disentangled version of the unitary operator (\ref{U}) is
\begin{equation}\label{FACTU}
U(\alpha,\tau) = e^{p_+ K_+} e^{p_0 K_0} e^{p_- K_-} \, ,
\end{equation}
where coefficients $p_+$, $p_0$, $p_-$, derived in \cite{TRUAX}, are known functions of $\alpha$ and $\tau$. Unlike in Sect. \ref{TRUAX}, two cases appear: $|\tau|^2 >$ $\alpha^2 / 4$ and $|\tau|^2 <$ $\alpha^2 / 4$. We set $\beta$ $\doteq$ $\left( \left|\, |\tau|^2 - \alpha^2 / 4 \,\right| \right)^{\frac{1}{2}}$. For $\displaystyle |\tau|^2 > \alpha^2 / 4$ one has
\begin{eqnarray}
p_+ &=& \frac{\tau \sinh \beta}{D} \, , \, p_- = - \frac{\bar{\tau}}{\tau} \, p_+ \, , \label{P+h}\\
p_0 &=& - 2 \ln \frac{D}{\beta} \, , \label{P0h}
\end{eqnarray}
where $D$ $=$ $\beta \cosh \beta$ $- i \displaystyle \frac{\alpha}{2} \sinh \beta$. Note that $p_-$ and $p_+$ differ from each other by a phase factor, so that in calcultions one can use $|p_-|$ $=$ $|p_+|$. When $\alpha$ $=$ 0 the disentanglement coefficients (\ref{T+h})-(\ref{T0h}) are naturally retrieved from (\ref{P+h})-(\ref{P0h}) as $U(0,\tau)$ $=$ $S(\tau)$. For $|\tau|^2 < \alpha^2 / 4$ the coefficients are derived from Eqs. (\ref{P+h})-(\ref{P0h}) by simply replacing the hyperbolic functions with the corresponding trigonometric ones. In the transition case $|\tau|^2$ $=$ $\alpha^2 / 4$, from Eqs. (\ref{P+h})-(\ref{P0h}) one obtains
\[
p_+ = \frac{2 \tau}{2 - i \alpha} \, , \, p_- = - \frac{\bar{\tau}}{\tau} p_+ \, , \, p_0 = - 2 \ln \left( 1 - i \frac{\alpha}{2} \right) \, .
\]
Note also that
\begin{equation}\label{PROPERTY}
e^{- \frac{1}{2} (p_0 + \bar{p}_0)} \left( 1 - |p_+|^2 \right) = 1 \, .
\end{equation}

Due to the Schwinger two boson realization, the generalized squeezed vacuum (\ref{GEN}) splits into the sum of two orthogonal definite-parity states,
\begin{equation}\label{GENKET}
|\alpha,\tau,\vartheta\rangle = \cos \vartheta \, |\alpha,\tau,0\rangle + \sin \vartheta \, |\alpha,\tau,\textstyle{\frac{\pi}{2}}\rangle \, ,
\end{equation}
where $|\alpha,\tau,0\rangle$ and $|\alpha,\tau,{\textstyle{\frac{\pi}{2}}}\rangle$ are obtained resorting to Eq. (\ref{FACTU}),
\begin{eqnarray*}
&&|\alpha,\tau,0\rangle = U(\alpha,\tau) |0\rangle \doteq \sum_{n=0}^{\infty} c_{2 n} \, |2 n\rangle \, , \\
&&|\alpha,\tau,\textstyle{\frac{\pi}{2}}\rangle = U(\alpha,\tau) |1\rangle \doteq \displaystyle \sum_{n=0}^{\infty} c_{2 n+1} \, |2 n +1\rangle \, ,
\end{eqnarray*}
with
\begin{eqnarray*}
c_{2 n} &=& e^{\frac{1}{4}p_0} \left(\frac{p_+}{2}\right)^n \frac{\sqrt{(2 n)!}}{n!} \, , \\
c_{2 n+1} &=& e^{\frac{3}{4}p_0} \left(\frac{p_+}{2}\right)^n \frac{\sqrt{(2 n+1)!}}{n!} \, .
\end{eqnarray*}
For $\alpha = 0$, $|\alpha,\tau,0\rangle$ returns $|\tau\rangle_0$. States $|\alpha,\tau,0\rangle$, $|\alpha,\tau,\frac{\pi}{2}\rangle$ are individually normalized by definition, as can be readily checked resorting to $\displaystyle \sum_{n=0}^{\infty} \left( \frac{y}{2} \right)^{2n} {2n\choose n}$ $=$ $(1-y^2)^{- \frac{1}{2}}$, with $y$ $=$ $|p_+|$, and to property (\ref{PROPERTY}). Calculation of the variances $\Delta^2(q )$ and $\Delta^2(p )$ in state (\ref{GENKET}) once more requires only a single transformation,
\[
U^\dagger (\alpha,\tau) a^\dagger U(\alpha,\tau) =  e^{- \frac{1}{2} p_0} \left( a^\dagger - p_- \, a \right) \, ,
\]
and its straightforward extensions to all relevant operators, so that the expectations are reconducted only to $|\omega\rangle$, $\langle \alpha,\tau,\vartheta| \bullet |\alpha, \tau, \vartheta\rangle$ $=$ $\langle \omega| U^\dagger(\alpha, \tau) \bullet U(\alpha, \tau)|\omega\rangle$. The calculation scheme is similar to, but of course more complicated than, the conventional case. One finds
\begin{eqnarray}
&& \Delta^2(q ) =  \frac{1}{2} + \left[ e^{- \frac{1}{2} (p_0 + \bar{p}_0)} |p_-|^2 - \frac{1}{2} \left( e^{- p_0 } p_- + \rm{c.c.} \right) \right] \nonumber\\
&+& \left[ e^{- \frac{1}{2} (p_0 + \bar{p}_0)} \left( 1 + |p_-|^2 \right) - \left(e^{- p_0 } p_-  + \rm{c.c.} \right) \right] \sin^2\!\vartheta \nonumber\\
&-& \frac{1}{2} \left[ e^{- \frac{1}{2} p_0} (1 - p_-) + \rm{c.c.} \right]^2 \sin^2\!\vartheta \, \cos^2\!\vartheta \, , \label{VARGq}
\end{eqnarray}
and
\begin{eqnarray}
&& \Delta^2(p ) = \frac{1}{2} + \left[ e^{- \frac{1}{2} (p_0 + \bar{p}_0)} |p_-|^2 + \frac{1}{2} \left( e^{- p_0 } p_- + \rm{c.c.} \right) \right] \nonumber\\
&+& \left[ e^{- \frac{1}{2} (p_0 + \bar{p}_0)} \left( 1 + |p_-|^2 \right) + \left(e^{- p_0 } p_-  + \rm{c.c.} \right) \right] \sin^2\!\vartheta \nonumber\\
&+& \frac{1}{2} \left[ e^{- \frac{1}{2} p_0} (1 - p_-) - \rm{c.c.} \right]^2 \sin^2\!\vartheta \, \cos^2\!\vartheta \, . \label{VARGp}
\end{eqnarray}
Previous results (\ref{VARq0}), (\ref{VARp0}) are retrieved from the general equations (\ref{VARGq}), (\ref{VARGp}) setting $\alpha$ $=$ 0 and $\vartheta$ $=$ 0. For $\alpha$ $=$ 0 and $\vartheta$ $=$ ${\textstyle{\frac{\pi}{2}}}$ one obtains the variances $\Delta^2(q )_1$ and $\Delta^2(p )_1$ in the state $|\tau\rangle_1$ $=$ $S(\tau) |1\rangle$, whose use in an interferometric setup can be found, e.g., in \cite{OLIVA},
\begin{eqnarray*}
\Delta^2(q )_1 &=& \frac{1}{2} + (1 + 3 \sinh^2|\tau|) + 3 \sinh |\tau| \cosh |\tau| \cos \Phi_\tau \, , \\
\Delta^2(p )_1 &=& \frac{1}{2} + (1 + 3 \sinh^2|\tau|) - 3 \sinh |\tau| \cosh |\tau| \cos \Phi_\tau \, .
\end{eqnarray*}

\section{Quadrature squeezing}\label{SQ}
\begin{figure}[htb]
   \centering
        \includegraphics[width=0.5\textwidth]{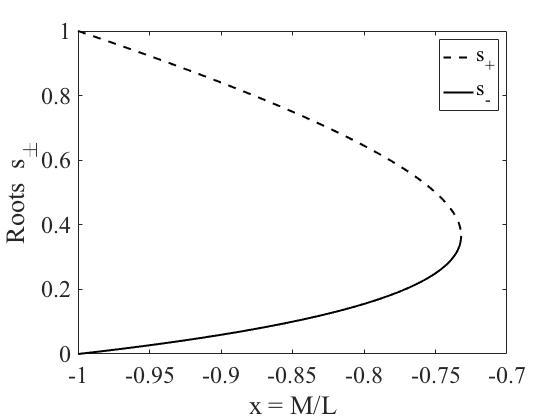}
    \caption{Squeezing transition curves corresponding to roots (\ref{SIN2FI}) of equation $\mathcal{G}(s)$ $=$ 0. $x$ is defined in $[-1, 1-\sqrt{3}]$.}
    \label{fig:S+-}
\end{figure}

Squeezing conditions for $q$ and $p$ are conveniently investigated writing the variances (\ref{VARGq}), (\ref{VARGp}) as polynomials in $s$ $\doteq$ $\sin^2\!\vartheta$, $0 \le s \le 1$. Setting
\begin{eqnarray}
\mathcal{F}  &=& \mathcal{F}(s) = A s^2 + B s + C \, , \label{FS} \\
\mathcal{G} &=& \mathcal{G}(s) = L s^2 + M s + N \, . \label{GS}
\end{eqnarray}
one obtains
\begin{equation}\label{VARGQP}
\Delta^2(q ) = \frac{1}{2} + \mathcal{F} - \mathcal{G} \; , \; \Delta^2(p ) = \frac{1}{2} + \mathcal{F} + \mathcal{G} \, .
\end{equation}
With $p_-$ $=$ $|p_-| e^{i \Phi_p}$, the explicit form of the (real) coefficients in Eqs. (\ref{FS}) and (\ref{GS}) in terms of the disentanglement coefficients (\ref{P+h}), (\ref{P0h}), is
\begin{eqnarray*}
A &=& e^{- \frac{1}{2} (p_0 + \bar{p}_0)} \left( 1 + |p_-|^2 - 2 |p_-| \cos \Phi_p \right) \\
B &=& e^{- \frac{1}{2} (p_0 + \bar{p}_0)} 2 |p_-| \cos \Phi_p \, , \\
C &=& e^{- \frac{1}{2} (p_0 + \bar{p}_0)} |p_-|^2 \, ,
\end{eqnarray*}
and
\begin{eqnarray*}
L &=& - \frac{1}{2} \left[ e^{- p_0 } \left( 1 - p_- \right)^2 + \rm{c.c.} \right] \, , \\
M &=& \frac{1}{2} \left[ e^{- p_0 } \left( 1 + p_-^2 \right) + \rm{c.c.} \right] \, , \\
N &=& \frac{1}{2} \left( e^{- p_0 } \, p_- + \rm{c.c.} \right) \, .
\end{eqnarray*}
Noting that $e^{- \frac{1}{2} (p_0 + \bar{p}_0)}$ is real and positive, $|p_-|^2 < 1$, and obviously $s^2 \le s$, $\mathcal{F}(s)$ proves to be positive for any $s$. In view of (\ref{VARGQP}) the squeezing transition between quadratures $q$ and $p$ corresponds to equation $\mathcal{G}(s)$ $=$ 0, whose roots, in terms of $x$ $\doteq$ $M / L$, are
\begin{equation}\label{SIN2FI}
s_\pm = \frac{1}{2} \left[ - x \pm \sqrt{x^2 - 2 x - 2} \right] \, .
\end{equation}
The range of values of $x$ in which constraint $0 \le s \le 1$ is satisfied is $[-1,1-\sqrt{3}]$. Roots $s_\pm$ in Fig. \ref{fig:S+-} show that for a given value of $x$ the squeezed quadrature is $q$ or $p$ according to whether $\mathcal{G}(s) > 0$, verified  for $0 < s < s_-$ and $s_ + < s < 1$, or $\mathcal{G}(s) < 0$, i.e., $s_- < s < s_+$, respectively. Note that $x$ $=$ $-1$ is the only condition for which the generalized vacuum $|\omega\rangle$ coincides with the ordinary vacuum $|0\rangle$ at $s_- = 0$ or with the single photon state $|1\rangle$ at $s_+ = 1$.

\section{Photon number statistics}\label{PHOTO}
\begin{figure}[htb]
    \centering
        \includegraphics[width=0.5\textwidth]{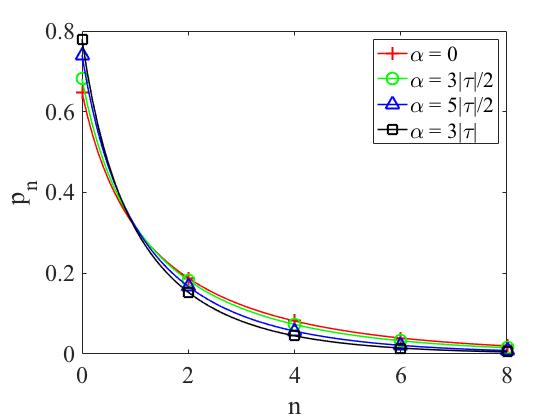}
    \caption{The even probability distribution (\ref{P2L}) for $|\tau|$ $=$ 1.}
    \label{fig:PNT1}
\end{figure}

\begin{figure}[htb]
    \centering
        \includegraphics[width=0.5\textwidth]{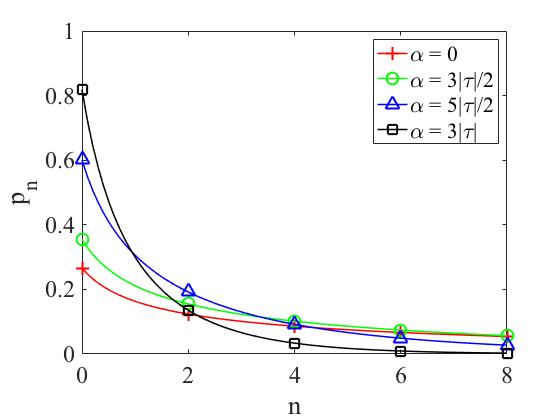}
    \caption{The even probability distribution (\ref{P2L}) for $|\tau|$ $=$ 2.}
    \label{fig:PNT2}
\end{figure}

\begin{figure}[htb]
    \centering
        \includegraphics[width=0.5\textwidth]{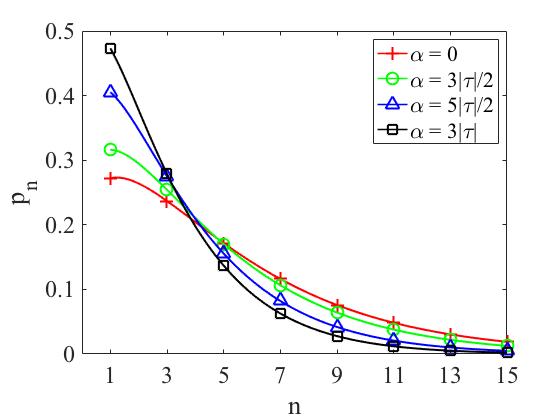}
    \caption{The odd probability distribution (\ref{P2L1}) for $|\tau|$ $=$ 1.}
    \label{fig:PNT1odd}
\end{figure}

\begin{figure}[htb]
    \centering
        \includegraphics[width=0.5\textwidth]{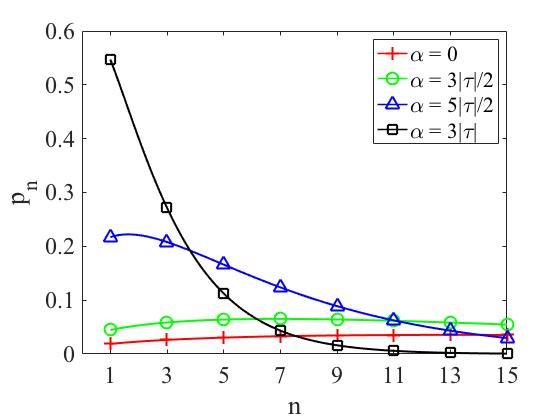}
    \caption{The odd probability distribution (\ref{P2L1}) for $|\tau|$ $=$ 2.}
    \label{fig:PNT2odd}
\end{figure}

A characteristic feature of the generalized squeezed vacuum $|\alpha, \tau, \vartheta\rangle$ is the photon number distribution $p_n$, that is the probability of finding $n$ photons in it. The even and odd photon number probability distributions, $p_{2 n}$ and $p_{2n+1}$, are
\begin{eqnarray}
p_{2 n} &=& |c_{2 n}|^2 = \left( 1 - |p_+|^2 \right)^{\frac{1}{2}} \left(\frac{|p_+|^2}{2^2}\right)^n {2n\choose n} \, , \label{P2L} \\
p_{2n+1} &=& |c_{2n+1}|^2 = \left( 1 - |p_+|^2 \right) (2n+1) \, p_{2n} \, , \label{P2L1}
\end{eqnarray}
where property (\ref{PROPERTY}) was used. Note that both distributions are separately normalized: $\sum_{n=0}^{\infty} p_{2 n}$ $=$ 1, $\sum_{n=0}^{\infty} p_{2 n + 1}$ $=$ 1. As expected Eq. (\ref{P2L}) gives the known result \cite{AGAR} for the conventional squeezed vacuum state $|\tau\rangle_0$ for $\alpha$ $=$ 0,
\[
{p_{2 n}} \Big|_ {(\alpha=0)} = \frac{1}{\cosh |\tau|} \left(\frac{\tanh^2|\tau|}{2^2}\right)^n {2n\choose n} \, .
\]
Probability distributions (\ref{P2L}) and (\ref{P2L1}) are shown in Figs. \ref{fig:PNT1}, \ref{fig:PNT2} and \ref{fig:PNT1odd}, \ref{fig:PNT2odd} (the continuous lines are drawn only for convenience). The cases $|\tau|$ $=$ 1, 2 are reported, while the control parameter $\alpha$ assumes the values $0$, $\frac{3}{2} |\tau|$,  $\frac{5}{2} |\tau|$, $3 |\tau|$, so that both conditions $|\tau|^2 > \alpha^2/4$ and $|\tau|^2 < \alpha^2/4$ occur. The probability distributions prove to be significantly influenced by $\alpha$. In particular, one notes that $p_{2n}$ and $p_{2n+1}$ are more sharply peaked on states $|0\rangle$ and $|1\rangle$ than in the reference case $\alpha$ $=$ 0, so that the spreading on the $n$ axis is generally reduced for both distributions. However, it is worth mentioning that, unlike $p_{2n}$, $p_{2n+1}$ can exhibit a maximum in $n$, which is not centered on state $|1\rangle$: in the explored range of values of $\alpha$ and $|\tau|$ this feature emerges in Fig. \ref{fig:PNT2odd} for $\alpha$ $=$ 0 and $3 |\tau|/2$, the maxima being very smooth. In the same range, numerical calculations show that as $|\tau|$ decreases both probability distributions $p_{2n}$ and $p_{2n+1}$ become practically independent on $\alpha$ for $|\tau| \le 0.4$.

Probability distributions $p_{2n}$, $p_{2n+1}$ can be both expressed in terms of the expectation values of the number operator in the even and odd parity states: $\langle \hat{n}\rangle_e$ $=$ $\sum_{n=0}^{\infty} (2n) \, p_{2 n}$ and $\langle \hat{n}\rangle_o$ $=$ $\sum_{n=0}^{\infty} (2n+1) \, p_{2 n + 1}$. Using (\ref{P2L}), (\ref{P2L1}) one calculates
\[
\langle \hat{n}\rangle_e = \frac{|p_+|^2}{1 - |p_+|^2} \; , \; \langle \hat{n}\rangle_o = \frac{1 + 2|p_+|^2}{1 - |p_+|^2} \, ,
\]
from which
\begin{eqnarray}
\!\! p_{2n} &=& ( \langle \hat{n}\rangle_e + 1)^{-\frac{1}{2}} \left[ \frac{\langle \hat{n}\rangle_e}{4 \, (\langle \hat{n}\rangle_e + 1)} \right]^n \, , \label{P2n<n>} \\
\!\! p_{2n+1} &=& 3^{\frac{3}{2}} (1+2n) ( \langle \hat{n}\rangle_o + 1)^{-\frac{3}{2}}\!\left[ \frac{\langle \hat{n}\rangle_o -1}{4 \, (\langle \hat{n}\rangle_o + 1)} \right]^n \!\!, \label{P2n1<n>}
\end{eqnarray}
manifestly different from Poisson's distribution. Note that, however, $p_{2n} $ and $p_{2n+1}$ depend only on $\langle \hat{n}\rangle_e$ and $\langle \hat{n}\rangle_o$, respectively, as it occurs, e.g., in Poissonian or thermal statistics. For generic $\vartheta$ the probability of finding $N$ photons in the squeezed vacuum $|\alpha,\tau,\vartheta\rangle$ is
\begin{equation}\label{PN}
p_N = | \langle N|\alpha,\tau,\vartheta\rangle |^2 = \cos^2\!\vartheta \, p_{2n} \, \delta_{N,2n} + \sin^2\!\vartheta \, p_{2n+1} \, \delta_{N,2n+1} \, ,
\end{equation}
as the interference terms vanish for any $N$ due to the form of Eq. (\ref{GENKET}). There ensues that the expectation value of number operator $\hat{n}$ in the generalized squeezed vacuum is
\[
\langle \hat{n} \rangle = \cos^2\!\vartheta \, \langle \hat{n} \rangle_e + \sin^2\!\vartheta \, \langle \hat{n} \rangle_o \, .
\]

\section{An application: the Mach-Zehnder interferometer}\label{MANDMZ}
\begin{figure}[!]
\centering
        \includegraphics[width=0.5\textwidth]{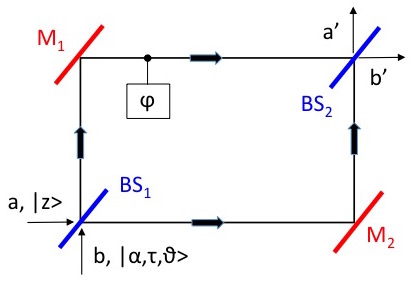}
    \caption{Schematic picture of the Mach-Zehnder interferometer: BS$_1$, BS$_2$ are 50-50 beam splitters, M$_1$, M$_2$ are perfect mirrors, $\varphi$ denotes the phase shift, here associated to the upper arm of the interferometer.}
    \label{fig:MZ}
\end{figure}

On account of the relevance of squeezed states for interferometry \cite{KENZIE}-\cite{PRADY}, as a paradigmatic example of an application of our generalized squeezed vacuum we consider the Mach-Zehnder interferometer described in Fig. \ref{fig:MZ}, where the conventions and nomenclature of \cite{AGAR} are adopted. The physical structure of the interferometer implies that one has to define two pairs of Heisenberg annihilation operators $(a,b)$ and $(a',b')$, at the input and output ports, respectively, which act on the corresponding Fock spaces here denoted by $\mathcal{F}_a $, $\mathcal{F}_b$ and $\mathcal{F}_{a'}$, $\mathcal{F}_{b'}$. These operators are related to each other through a unitary transformation expressed by the 2$\times$2 matrix $\mathbf{T}$,
\[
\left[ \begin{array}{l} a' \\ b' \end{array} \right] = \mathbf{T} \left[ \begin{array}{l} a \\ b \end{array} \right] = \left[ \begin{array}{c c } T_{11} & T_{12} \\ T_{21} & T_{22} \\ \end{array} \right] \left[ \begin{array}{l} a \\ b \end{array} \right] \, ,
\]
whose elements are $T_{11}$ $=$ $- T_{22}$ $=$ $- \frac{1}{2} \left( 1-e^{-i \varphi} \right)$, $T_{21}$ $=$ $T_{12}$ $=$ $- \frac{i}{2} \left( 1+e^{-i \varphi} \right)$. In tensor product notation the Hilbert space of the system is $\mathcal{H}$ $=$ $\mathcal{F}_a \otimes \mathcal{F}_b$ at the input ports and $\mathcal{H}'$ $=$ $\mathcal{F}_{a'} \otimes \mathcal{F}_{b'}$ $\sim$ $\mathcal{H}$ at the output ports. With no risk of ambiguity in Fig. \ref{fig:MZ} we identify the ports labels with the same symbols of the corresponding annihilation operators.

We analyze the physical process in which the generalized squeezed vacuum state $|\alpha,\tau,\vartheta\rangle$ enters input port $b$, while the Glauber coherent state $|z\rangle$, $z \in \mathbb{C}$, enters port $a$. Let us recall that, defining the unitary operator $D(z)$ $=$ $\exp(z a^{\dagger} - \bar{z} a)$, $|z\rangle$ $=$ $D(z) |0\rangle$. It is worth noting that also $|z\rangle$ is an example of displaced vacuum belonging to the family of generalized coherent states defined in Section \ref{INTRO}, its algebra $A$ being the Heisenberg-Weyl algebra generated by the set of operators $\{\mathbb{I}, a, a^\dagger\}$, while $|\omega\rangle$ coincides with the physical vacuum $|0\rangle$ annihilated by $a$. Moreover, disentangling $D(z)$ in normal order form, $D(z)$ $=$ $e^{-{\frac{1}{2}} |z|^2} e^{z a^{\dagger}} e^{- \bar{z} a}$, one can write
\[
|z\rangle = e^{-\frac{1}{2} |z|^2} e^{z a^{\dagger}} |0\rangle = e^{-\frac{1}{2} |z|^2} \sum_{n=0}^{\infty} \frac{z^n}{\sqrt{n!}} |n\rangle \, .
\]
We explore the possibility of extended and controllable quantumness domains at the output of such process through the Mandel parameter $Q$ \cite{QM},
\begin{equation}\label{Q}
Q = \frac{\Delta^2(\hat{n}) - \langle \hat{n} \rangle}{\langle \hat{n} \rangle} = \frac{\langle \hat{n}^2 \rangle}{\langle \hat{n} \rangle} - \langle \hat{n} \rangle - 1 \, ,
\end{equation}
as it is known that the sign of $Q$ classifies the statistics of the photon number distribution: negative (positive) values of $Q$ correspond to states with respect to which this statistics is sub (super)-Poissonian ($Q = 0$ identifies Poisson's statistics). Using the additional control parameters $\alpha$, $\vartheta$, that the generalized squeezed vacua make available, one shows the existence of new nonclassical regions with respect to the standard scheme in which $\alpha$ $=$ 0, $\vartheta$ $=$ 0 and the state $|\alpha,\tau,\vartheta\rangle$ at the input port $b$ reduces to $|\tau\rangle_0$. With our input state $|z\rangle \otimes |\alpha,\tau,\vartheta\rangle$, we study the output at port $a'$; therefore $a'$ $=$ $T_{11} \, a \otimes \mathbb{I} + T_{12} \, \mathbb{I} \otimes b$ and the number operator $\hat{n}$ in Eq. (\ref{Q}) is identified with $\hat{n}_{a'}$ $\doteq$ $a'^\dagger a'$,
\begin{equation}\label{NOUT}
\hat{n}_{a'} = |T_{11}|^2 \hat{n}_a \otimes \mathbb{I} + |T_{12}|^2 \mathbb{I} \otimes \hat{n}_b + \left[ \bar{T}_{11} T_{12} a^\dagger \otimes b + T_{11} \bar{T}_{12} a \otimes b^\dagger \right] \, ,
\end{equation}
where $\hat{n}_a$ $\doteq$ $a^\dagger a$, $\hat{n}_b$ $\doteq$ $b^\dagger b$. Using Eq. (\ref{NOUT}) and formula (\ref{PROPERTY}), bearing in mind that $|p_+|^2$ $=$ $|p_-|^2$, the expectation values of $\hat{n}_{a'}$ and $\hat{n}^2_{a'}$ for the configuration in Fig. \ref{fig:MZ} are, explicitly,

\begin{eqnarray*}
&&\langle \hat{n}_{a'} \rangle = \langle \omega| U^\dagger(\alpha,\tau) \langle z| \hat{n}_{a'} |z\rangle U (\alpha,\tau) |\omega\rangle = |z|^2 \sin^2\!\frac{\varphi}{2}  \\
&& + \left( 1 - |p_-|^2 \right)^{-1} \left[ |p_-|^2 + \left( 1 + |p_-|^2 \right) \sin^2\!\vartheta \right] \cos^2\!\frac{\varphi}{2} \\
&& + \frac{1}{4} \left[ z e^{- \frac{1}{2} p_0} (1-p_-) + \textrm{c.c.} \right] \sin(2 \vartheta) \sin \varphi \, ,
\end{eqnarray*}

\begin{eqnarray*}
&&\langle \hat{n}^2_{a'} \rangle = \langle \omega| U^\dagger(\alpha,\tau) \langle z| \hat{n}^2_{a'} |z\rangle U (\alpha,\tau) |\omega\rangle = \sin^4\!\frac{\varphi}{2} \, |z|^2 \left( 1 + |z|^2 \right) \\
&& + \cos^4\!\frac{\varphi}{2} \,\, \left( 1 - |p_-|^2 \right)^{-2} \left[ \left( 1+8|p_-|^2+3|p_-|^4 \right) \sin^2\!\vartheta + |p_-|^2 \left(2+|p_-|^2\right) \right] \\
&&+ \sin^2\!\varphi \,  \left( 1 - |p_-|^2 \right)^{-1} \left[ |p_-|^2 + \left( 1 + |p_-|^2 \right) \sin^2\!\vartheta \right] \left( \frac{1}{4} + |z|^2 \right) \\
&& + \frac{1}{4} \sin^2\!\varphi \left[ |z|^2 - \left(  z^2 \, e^{-p_0} p_- + \textrm{c.c.} \right) \left( 1 + 2 \sin^2\!\vartheta \right) \right] \\
&&+ \frac{1}{4} \sin(2\vartheta) \sin \varphi \left\{ \left( 1 + 2 |z|^2 \right) \left[ z \, e^{- \frac{1}{2} p_0} ( 1 - p_- ) + \textrm{c.c.} \right] \sin^2\!\frac{\varphi}{2} \right.\\
&&\left. + \left[ z \, e^{- \frac{1}{2} p_0} \left( 1 - |p_-|^2 \right)^{-1} \left[ 1 + 5 |p_-|^2 - 3 p_- (1+|p_-|^2) \right] + \textrm{c.c.} \right] \cos^2\!\frac{\varphi}{2} \right\} \, .
\end{eqnarray*}
\noindent\\
Selecting port $b'$ would lead to different expectation values, which are readily computable noticing that, with $\hat{n}_{b'}$ $\doteq$ $b'^\dagger b'$, operators $\hat{n}_{b'}$ and $\hat{n}^2_{b'}$ are obtained from $\hat{n}_{a'}$ and $\hat{n}^2_{a'}$, respectively, replacing $\varphi$ with $\varphi \pm \pi$ throughout.

\begin{figure}[htb]
\centering
        \includegraphics[width=0.5\textwidth]{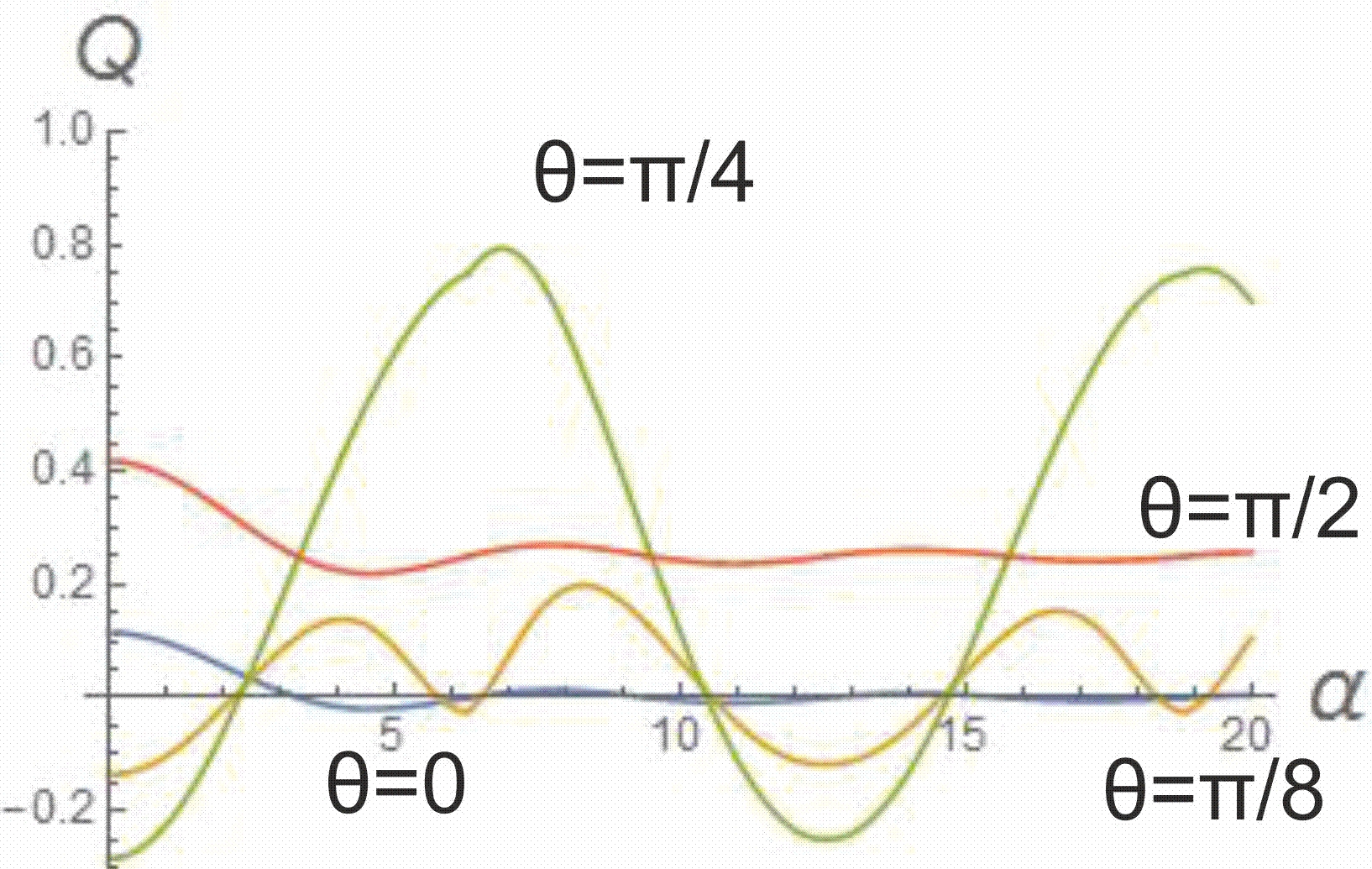}
    \caption{Mandel's parameter as a function of $\alpha$ for $\tau$ $=$ 0.1. The blue, yellow, green and red curves are labelled $\vartheta$ $=$ $0$, $\frac{\pi}{8}$, $\frac{\pi}{4}$ and $\frac{\pi}{2}$.}
    \label{fig:Q1}
\end{figure}

\begin{figure}[htb]
\centering
        \includegraphics[width=0.5\textwidth]{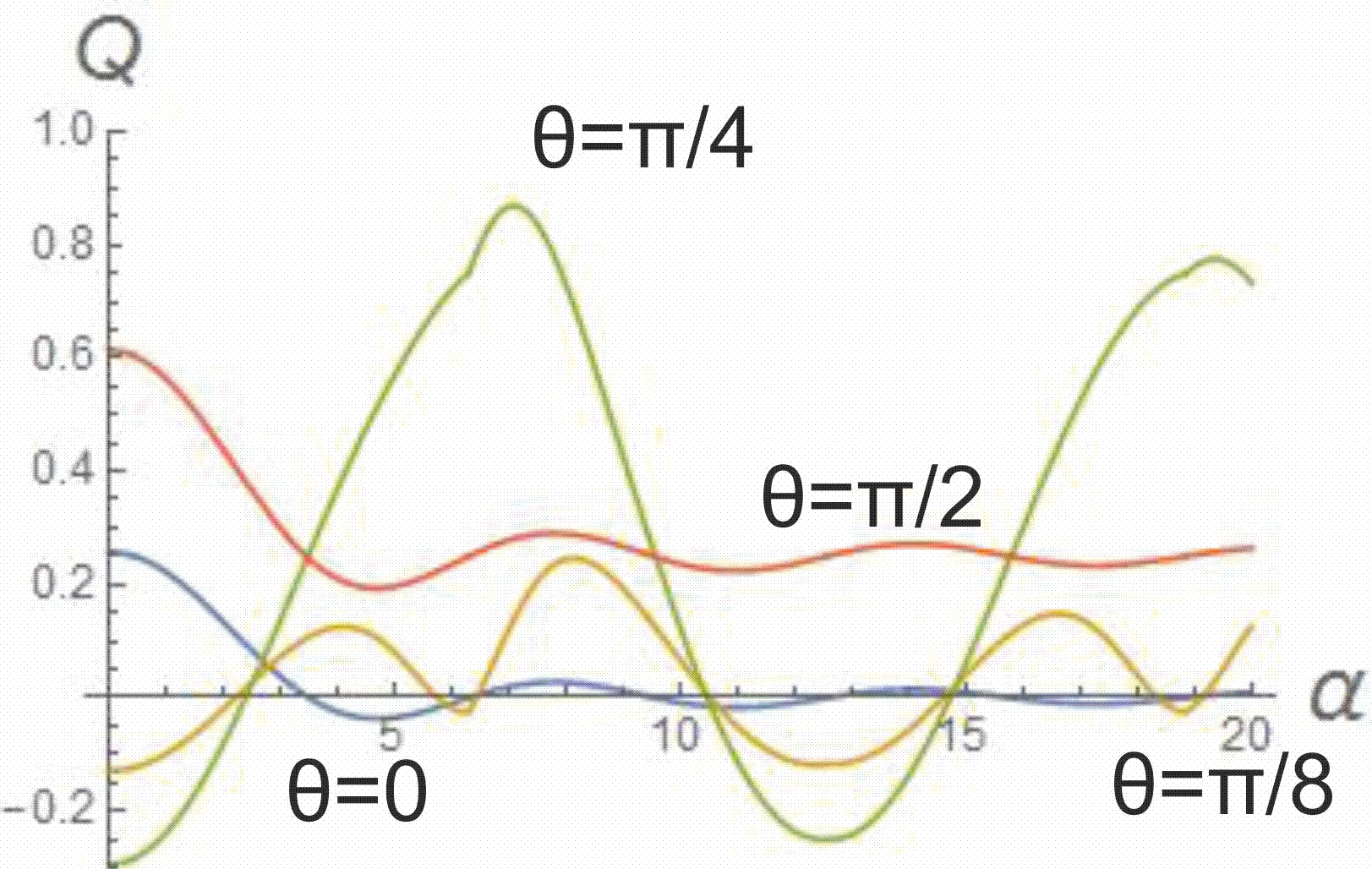}
    \caption{Mandel's parameter as a function of $\alpha$ for $\tau$ $=$ 0.2. The blue, yellow, green and red curves are labelled $\vartheta$ $=$ $0$, $\frac{\pi}{8}$, $\frac{\pi}{4}$ and $\frac{\pi}{2}$.}
    \label{fig:Q2}
\end{figure}

\begin{figure}[htb]
\centering
        \includegraphics[width=0.5\textwidth]{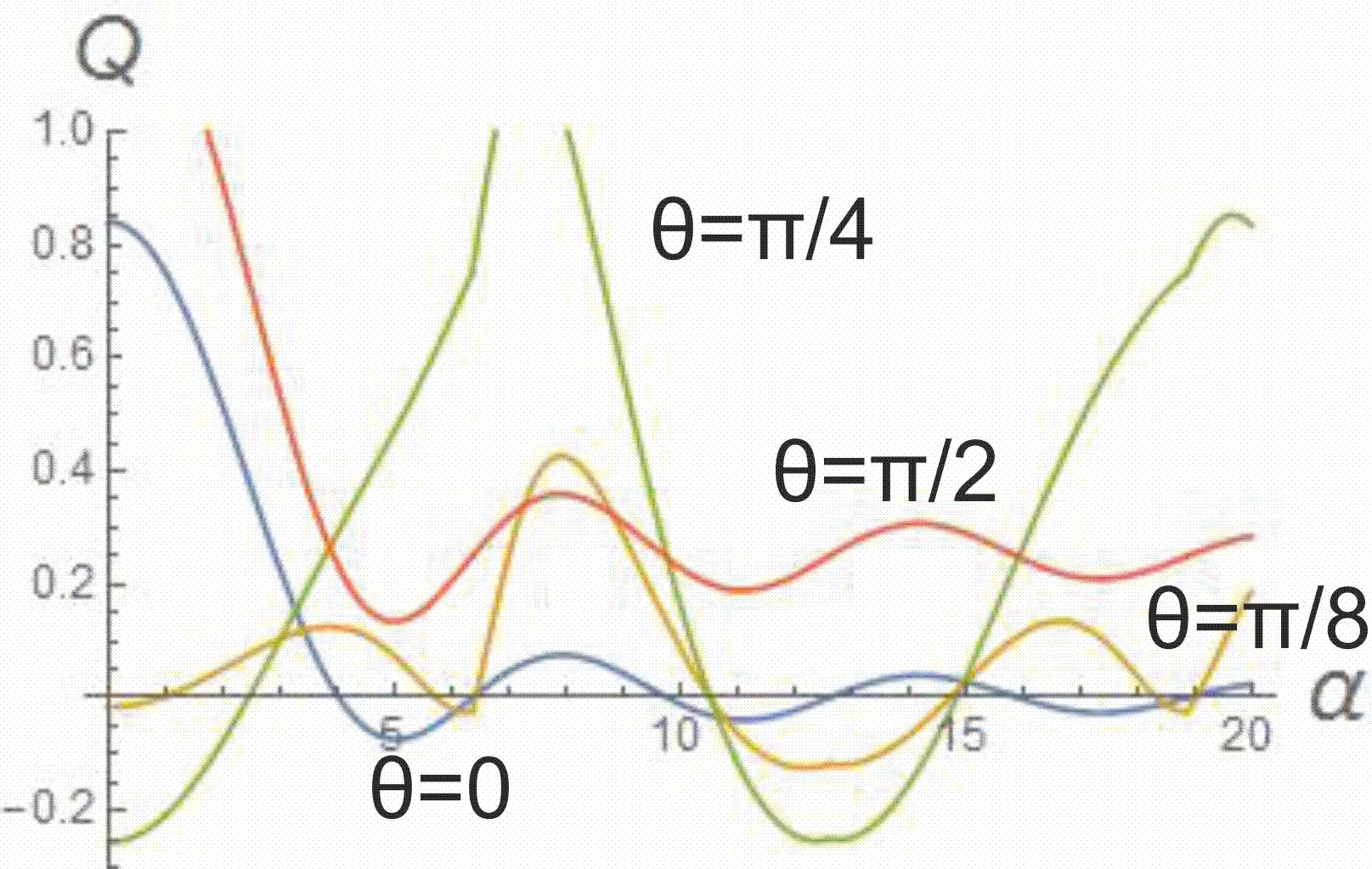}
    \caption{Mandel's parameter as a function of $\alpha$ for $\tau$ $=$ 0.5. The blue, yellow, green and red curves are labelled $\vartheta$ $=$ $0$, $\frac{\pi}{8}$, $\frac{\pi}{4}$ and $\frac{\pi}{2}$.}
    \label{fig:Q3}
\end{figure}

\begin{figure}[htb]
\centering
        \includegraphics[width=0.5\textwidth]{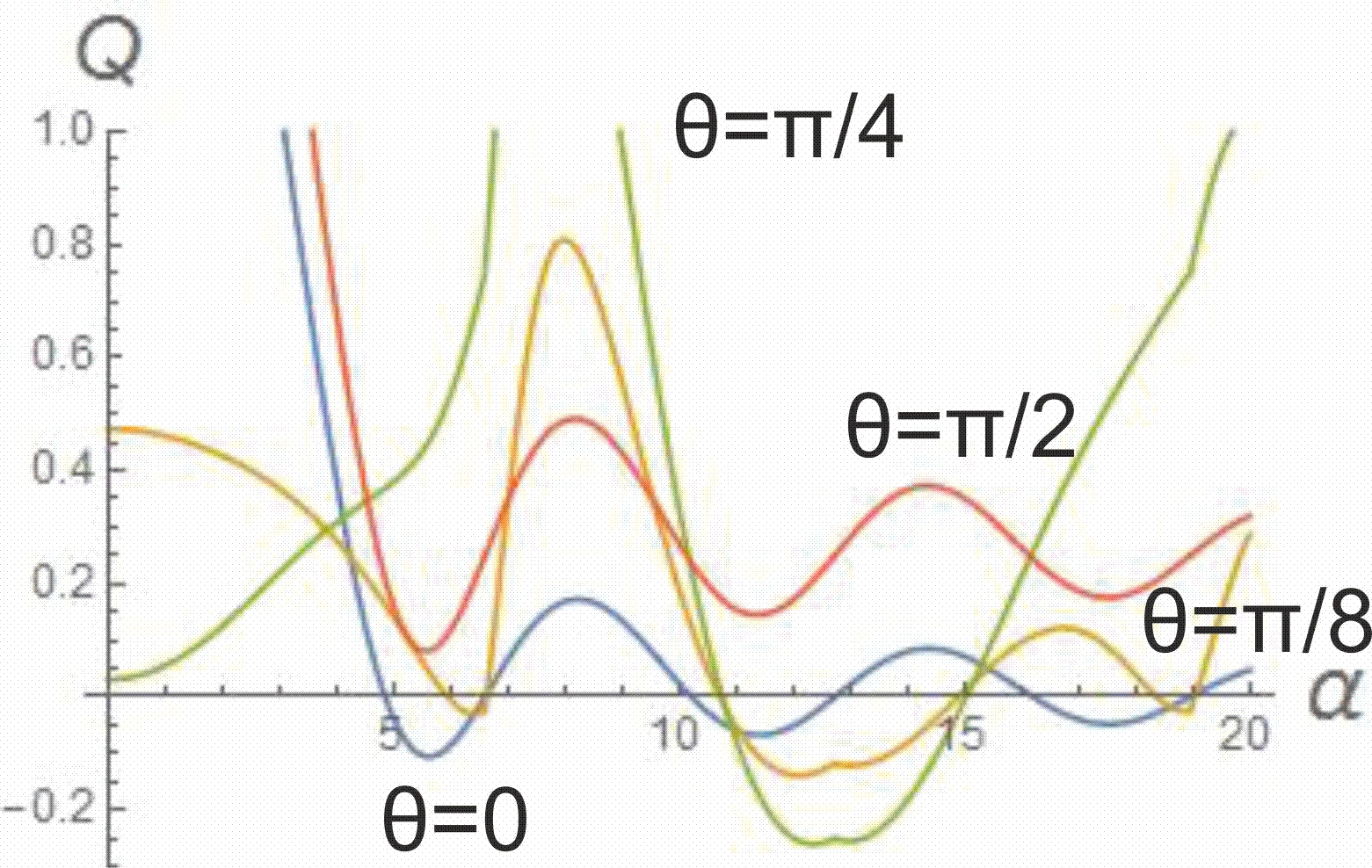}
    \caption{Mandel's parameter as a function of $\alpha$ for $\tau$ $=$ 1. The blue, yellow, green and red curves are labelled $\vartheta$ $=$ $0$, $\frac{\pi}{8}$, $\frac{\pi}{4}$ and $\frac{\pi}{2}$.}
    \label{fig:Q4}
\end{figure}

In our example we fix $z$ $=$ 1, $\varphi$ $=$ $\frac{\pi}{2}$ (selecting other values for $z$ and $\varphi$ would simply lead to other sub-Poissonian regions in the parameter space) and calculate the Mandel parameter (\ref{Q}) for various values of both $\tau$ and the additional degrees of freedom $\alpha$ and $\vartheta$. Each curve in Figs. \ref{fig:Q1}-\ref{fig:Q4} corresponds to a different value of $\vartheta$ while $\tau$ is different in every figure. One can see that nonclassical domains, characterized by negative values of $Q$, emerge even for $\vartheta$ $=$ 0 ($|\omega\rangle$ $\equiv$ $|0\rangle$) in some intervals of the axis $\alpha$. Conversely, nonclassical behaviour arises for $\alpha$ $=$ 0, where $U(0,\tau)$ $\equiv$ $S(\tau)$, depending on the value of $\vartheta$. The numerical results show that higher values of $\tau$ counteract this phenomenon, namely, the nonclassical regions are removed or shifted toward higher values of $\alpha$, depending on the interplay among the parameters. The tiny breaks of smoothness in Figs. \ref{fig:Q1}-\ref{fig:Q4} happen in the region $|\tau|^2 < \alpha^2/4$ for $\alpha^2/4 - |\tau|^2$ $=$ $(k \pi)^2$, $k \ne 0$ integer, where $\sin \beta$ $=$ 0 or, equivalently, $|p_-|$ $=$ 0.

\begin{figure}[htb]
\centering
        \includegraphics[width=0.5\textwidth]{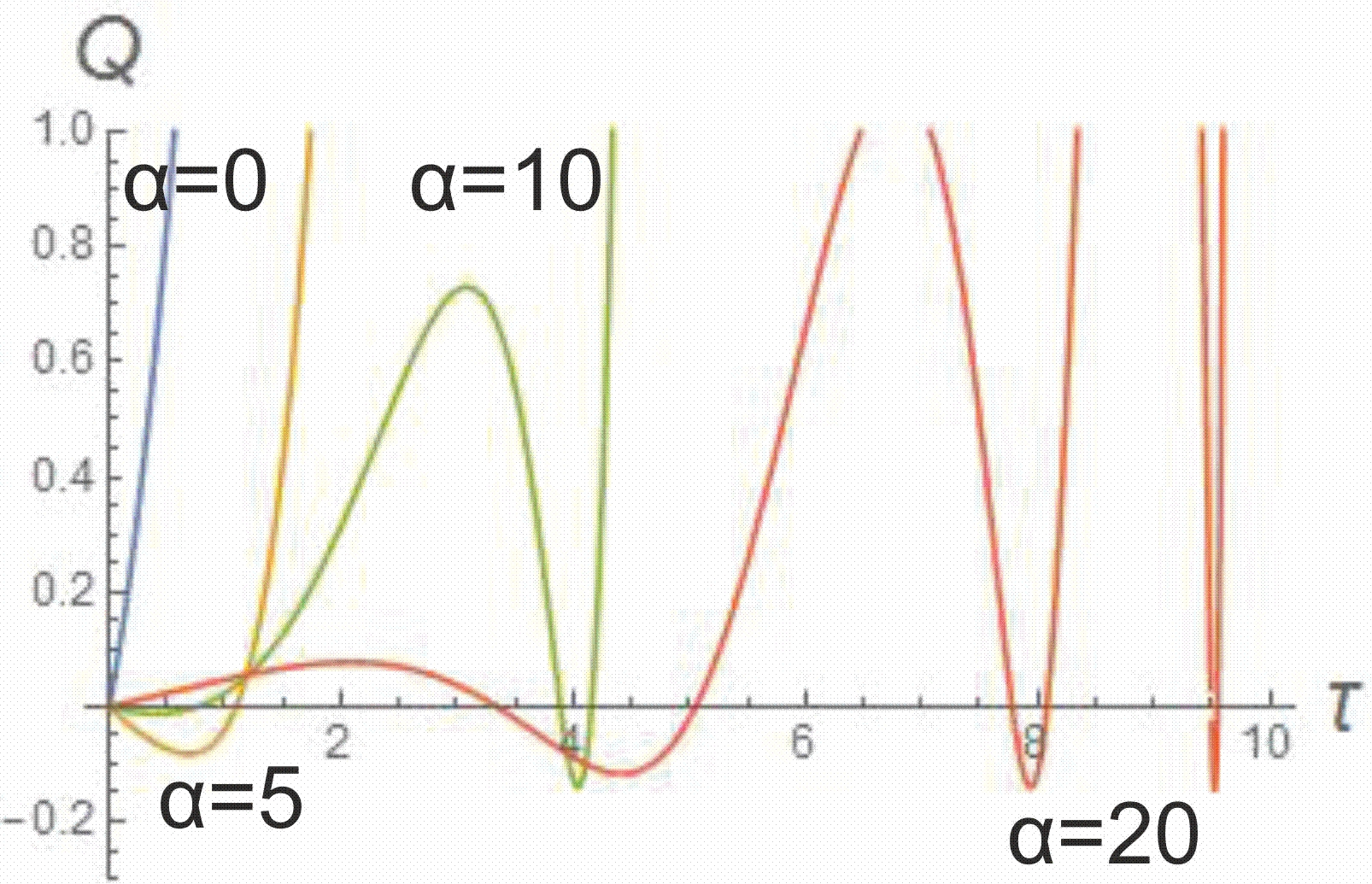}
    \caption{Mandel's parameter as a function of $\tau$ for $\vartheta$ $=$ 0. The blu, yellow, green and red curves are labelled $\alpha$ $=$ 0, 5, 10 and 20.}
    \label{fig:Q5}
\end{figure}

\begin{figure}[htb]
\centering
        \includegraphics[width=0.5\textwidth]{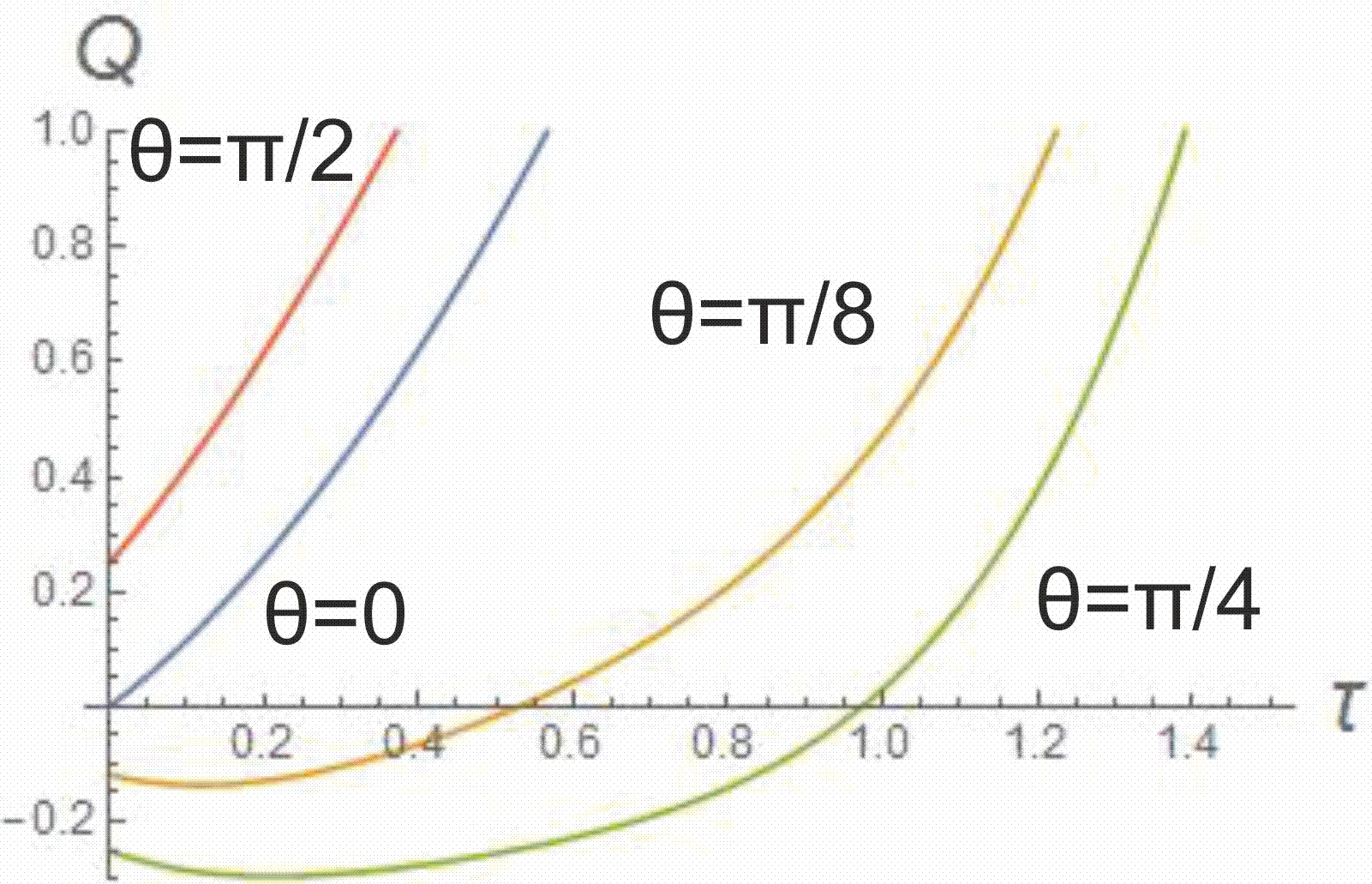}
    \caption{Mandel's parameter as a function of $\tau$ for $\alpha$ $=$ 0. The blu, yellow, green and red curves are labelled $\vartheta$ $=$ 0, $\frac{\pi}{8}$, $\frac{\pi}{4}$ and $\frac{\pi}{2}$.}
    \label{fig:Q6}
\end{figure}

\begin{figure}[htb]
\centering
        \includegraphics[width=0.5\textwidth]{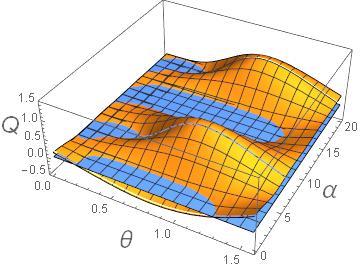}
    \caption{Mach-Zehnder example: $Q(\alpha, \vartheta)$ for $\tau$ $=$ 0.1.}
    \label{fig:Q9}
\end{figure}

\begin{figure}[htb]
\centering
        \includegraphics[width=0.5\textwidth]{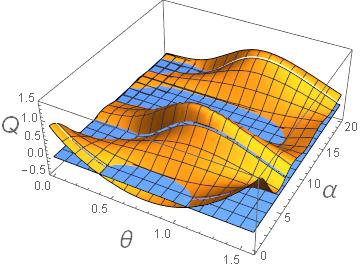}
    \caption{Mach-Zehnder example: $Q(\alpha, \vartheta)$ for $\tau$ $=$ 0.5.}
    \label{fig:Q10}
\end{figure}

\begin{figure}[htb]
\centering
        \includegraphics[width=0.5\textwidth]{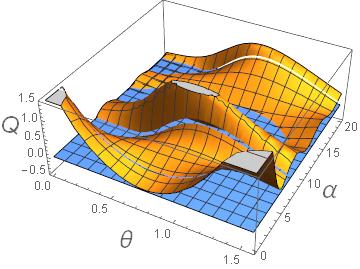}
    \caption{Mach-Zehnder example: $Q(\alpha, \vartheta)$ for $\tau$ $=$ 1.}
    \label{fig:Q11}
\end{figure}

\begin{figure}[htb]
\centering
        \includegraphics[width=0.5\textwidth]{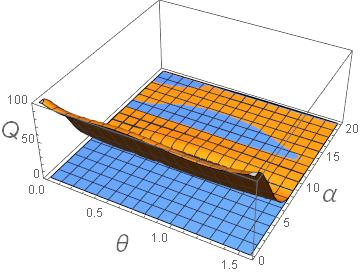}
    \caption{Mach-Zehnder example: $Q(\alpha, \vartheta)$ for $\tau$ $=$ 3.}
    \label{fig:Q12}
\end{figure}

Figs. \ref{fig:Q5}-\ref{fig:Q6} show $Q$ as a function of $\tau$ considering separately the additional degrees of freedom $\alpha$ and $\vartheta$: in Fig. \ref{fig:Q5} $\vartheta$ $=$ 0 and the curves are labelled by $\alpha$ whereas in Fig. \ref{fig:Q6} $\alpha$ $=$ 0 and the curves are parametrized by $\vartheta$. The emergence of sub-Poissonian regimes is visually noticeable with respect to the usual setting, which corresponds to both $\alpha$ $=$ 0 and $\vartheta$ $=$ 0. Finally, the 3D plots in Figs. \ref{fig:Q9}-\ref{fig:Q12} give the joint dependence of the Mandel parameter on the control degrees of freedom $\alpha$ and $\vartheta$. In each figure the surface $Q(\alpha, \vartheta)$ is associated to a specific value of $\tau$ and the blue plane denotes $Q$ $=$ 0. Increasing $\tau$ reduces the extension of the sub-Poissonian domains; this effect is emphasized in Fig. \ref{fig:Q12} where $\tau$ attains the maximum value we assumed for this example. As a general consideration, these results show that a rich variety of regions of non classical behaviour appears: this paves the way to possible interesting applications of these states to interferometry.



\section{Discussion and final remarks}\label{END}
In this paper a twofold generalization of the usual approach to quantum squeezed states was presented, which resorts to a new set of squeezed states based. The latter are based  on the rigorous mathematical definition of coherent states for the two boson Schwinger realization of the algebra su(1,1). Such formulation leads to generalized expressions of quadratures squeezing and photon probability distribution, which include the standard results as special cases. Furthermore, new nonclassical regions are predicted in the application to a Mach-Zehnder interferometer. Specifically:
\begin{itemize}
\item State $|\alpha,\tau,\vartheta\rangle$, Eq.  (\ref{GEN}), is the coherent state for su(1,1) and provides a generalized squeezed vacuum state, which includes the conventional squeezed vacuum state  $|\tau\rangle_0$. Compared to $|\tau\rangle_0$, $|\alpha,\tau,\vartheta\rangle$ exhibits two features: it lives in the full Fock space and not in the even sector only and gives to the experimenter two additional degrees of freedom, $\alpha$ and $\vartheta$, which could be used as control parameters of an actual quantum optical system to explore quantumness in a larger parameter space.
\item The generalized formulation leads to the identification of which is the squeezed quadrature, cf. Fig. \ref{fig:S+-}. The regions accessible to the system can be selected by tuning a suitable parameter $x$, which depends on coefficients $\alpha$ and $\tau$. By comparison, the conventional case corresponds to the single point $(-1,0)$ in Fig. \ref{fig:S+-}.
\item The influence of the control parameter $\alpha$ on photon statistics is significant. This is shown in Figs. \ref{fig:PNT1}, \ref{fig:PNT2} and \ref{fig:PNT1odd}, \ref{fig:PNT2odd} for the special cases $\vartheta$ $=$ 0 and $\vartheta$ $=$ $\pi/2$, i.e., $|\omega\rangle$ $=$ $|0\rangle$ and $|\omega\rangle$ $=$ $|1\rangle$, respectively.
\item In the Mach-Zehnder example, with $\varphi$ $=$ $\frac{\pi}{2}$, $z$ $=$ 1, Mandel's parameter exhibits a marked dependence on the control parameters $\alpha$, $\vartheta$. Indeed, acting on $\alpha$ and $\vartheta$ gives rise to nonclassical regions in which the statistics is sub-Poissonian. This is  illustrated in Figs. \ref{fig:Q1}-\ref{fig:Q4}. The curves in Figs. \ref{fig:Q5} and \ref{fig:Q6}, which provide $Q$ as a function of $\alpha$ for $\vartheta$ $=$ 0 and of $\vartheta$ for $\alpha$ $=$ 0, allow for a straightforward comparison with the standard case ($\alpha = 0$, $\vartheta = 0$). In all cases considered parameter $\tau$, which is assumed to be real in the numerical calculations, plays an important role as well: the higher its value, the less pronounced the nonclassical effects induced by the control parameters. This is confirmed by the three-dimensional plots in Figs. \ref{fig:Q9}-\ref{fig:Q12}, which  summarize the previous considerations and show the influence of the interplay between the control parameters $\alpha$, $\vartheta$ and $\tau$ on the extension of the nonclassical regions of the system. The presence of several sub-Poissonian domains motivates the potential experimental interest of the quantum squeezing formulation proposed in this work.
\end{itemize}

\section{Acknowledgments}
One of the authors (MG) acknowledges the support of  the grant 190492 of FQXI foundation.

\section*{References}

\end{document}